\documentclass[prb,reprint,superscriptaddress,super,superbib,amsmath]{revtex4-1}
\usepackage{graphicx}
\usepackage[squaren]{SIunits}

\begin{document}

\title{Ultrafast spectral diffusion measurement on nitrogen vacancy centers in nanodiamonds
using correlation interferometry}

\author{Janik Wolters} \email[Electronic mail: ]{janik.wolters@physik.hu-berlin.de}
\affiliation{Nano-Optics, Institute of Physics, Humboldt-Universit\"{a}t zu
Berlin, Newtonstr.~15, D-12489  Berlin, Germany}

\author{Nikola Sadzak}
\affiliation{Nano-Optics, Institute of Physics, Humboldt-Universit\"{a}t zu
Berlin, Newtonstr.~15, D-12489  Berlin, Germany}

\author{Andreas W. Schell}
\affiliation{Nano-Optics, Institute of Physics, Humboldt-Universit\"{a}t zu
Berlin, Newtonstr.~15, D-12489  Berlin, Germany}

\author{Tim Schr\"oder}
\affiliation{Nano-Optics, Institute of Physics, Humboldt-Universit\"{a}t zu
Berlin, Newtonstr.~15, D-12489  Berlin, Germany}

\author{Oliver Benson}
\affiliation{Nano-Optics, Institute of Physics, Humboldt-Universit\"{a}t zu
Berlin, Newtonstr.~15, D-12489  Berlin, Germany}

\begin{abstract}
Spectral diffusion is the phenomenon of random jumps in the emission wavelength of narrow lines. This phenomenon
is a major hurdle for applications of solid state quantum emitters like quantum dots, molecules or diamond
defect centers in an integrated quantum optical technology. Here, we provide further insight into the underlying
processes of spectral diffusion of the zero phonon line of single nitrogen vacancy centers in nanodiamonds by
using a novel method based on photon correlation interferometry. The method works although the spectral
diffusion rate is several orders of magnitude higher than the photon detection rate and thereby improves the
time resolution of previous experiments with nanodiamonds by six orders of magnitude. We study the dependency of
the spectral diffusion rate on the excitation power, temperature, and excitation wavelength under off-resonant
excitation. Our results bring insight into the mechanism of spectral diffusion and  suggest a strategy to increase the number of spectrally indistinguishable photons
emitted by diamond nanocrystals.
\end{abstract}

\maketitle

In the last decade the negatively charged nitrogen vacancy center (NV) in diamond has emerged as a promising
resource for future quantum technology~\cite{Ladd2010,Hanson2008,Brien2009,Aharonovich2011,Jelezko2006}. NVs,
consisting of a nitrogen impurity atom with an adjacent vacancy, occur naturally in bulk diamond or diamond
nanocrystals and are commonly used as single photon sources, operating even at
room-temperature~\cite{Wrachtrup2006}. Diamond nanocrystals with a single NV are of particular interest,
as they can be deterministically integrated into photonic and plasmonic hybrid devices and
circuits~\cite{Wolters2010, Wolters2011,VanderSar2011,Benson2011,Englund2010}. The spectrum of the NV
single photon emission consist of a broad phonon sideband around 700~nm and a narrow zero phonon line (ZPL) at
about 638 nm. 
At cryogenic temperatures the ZPL of single NVs in bulk type IIa diamond has been shown to be
almost Fourier-limited under off-resonant excitation, which allowed for the observation of two-photon
interference ~\cite{Bernien2011,Sipahigil2011}, a crucial step towards applications in optical quantum
information processing. 
However, for implanted NVs located close to the surface, several GHz wide spectral diffusion can be observed under off-resonant illumination.\cite{Fu2010} 
In nitrogen rich type Ib nanodiamonds the energy levels fluctuate much stronger, leading to a broadened ZPL with a width of about 0.5~nm. It is widely assumed, that line broadening of
emitters in condensed phase is due to a fluctuating electrostatic environment~\cite{Majumdar2011,Zander2002}. In
diamond nanocrystals it is caused by ionized impurities \cite{Farrer1969, Rosa1999} and charge traps.
The spectral shift caused by single charges can be estimated using a simple toy model
(sketched in Fig.~1a). Using data from Refs.~\cite{Acosta2012, Tamarat2006} we found that the
Stark-shift of the excited states $\Delta \mathcal{E}_{x/y}=d\cdot E$ in an NV center induced by a single elementary charge in a distance of about 10~nm can be as large as several hundred GHz.
The time  after which the probability that the electrostatic environment remains unchanged is reduced to $1/e$
can be defined as the spectral diffusion time $\tau_{D}$. Within $\tau_{D}$  the system can be assumed to be free from
spectral diffusion, and all photons emitted within $\tau_{D}$ should be nearly indistinguishable. Enhancing the
number of indistinguishable subsequent photons is a key goal.
In this paper, we present measurements of  $\tau_{D}$ in NV centers in nanodiamonds under off-resonant excitation with a time-resolution exceeding previous
experiments~\cite{Shen2008} with nanodiamonds by about six orders of magnitude.
Our studies identify the excitation laser as the main source for enhanced spectral diffusion.\\
\begin{figure}[b]
\centering
 \includegraphics[width=0.9\columnwidth]{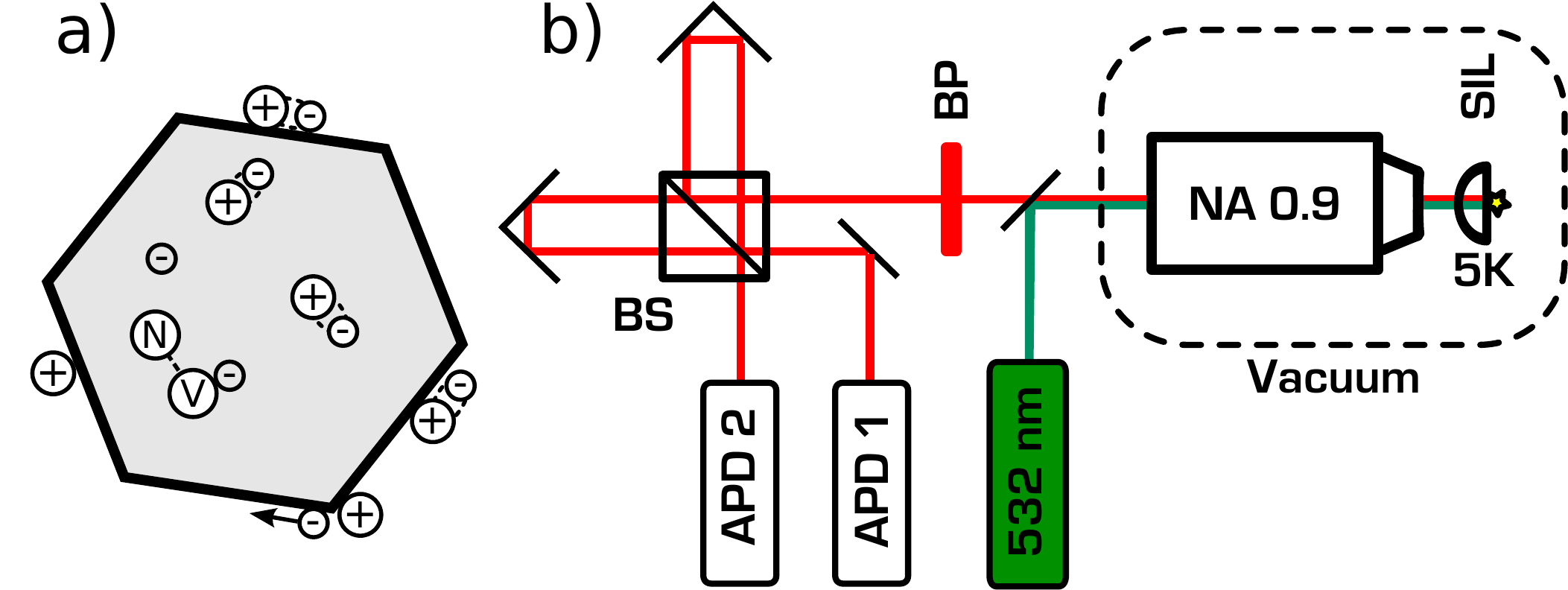}
  \caption{(Color online) a) Toy model for spectral diffusion. Charge traps are ionized and lead to fluctuating electric fields at the NV center's location.
  b) Sketch of the experimental setup. Milled type Ib nanodiamonds are deposited on a solid immersion lens (SIL) and placed in a continuos flow \mbox{He-cryostat}.
Single photon fluorescence is excited with a 532~nm laser and collected through an objective lens with numerical
aperture NA = 0.9. Using a spectral filter (BP) photons from the zero-phonon transition are filtered and sent
through a folded Mach-Zehnder interferometer. }
  \label{fig:2}
\end{figure}
The most straightforward method to measure spectral diffusion in single emitters is to take a time series of
spectra and directly visualize the spectral wandering~\cite{Zander2002,Bimberg2008,Sallen2010}. 
However, this method is only suitable for low spectral diffusion rates.
For many emitters, like NV centers in nanodiamonds the rate of detected photons emitted from the ZPL transition is
on the order a few kHZ, while spectral diffusion occurs much faster.\\
In the literature two methods which in principle enable spectral diffusion measurements on timescales as fast as the
emitter's lifetime were suggested and demonstrated recently~\cite{Brokmann2006,Coolen2008}. The idea is to
convert the frequency fluctuations due to spectral diffusion into intensity fluctuations. However, in this
approach, technical limitations such as the required complex setup with several moving parts reduced the time
resolution in the first measurements to the order of 100 $\mu$s~\cite{Coolen2008}. In a related approach Sallen et al.~\cite{Sallen2010}
reached the subnanosecond regime, but for the price of a spectrometer-limited spectral resolution and an
inherently reduced detection efficiency. As pointed out in Ref.~\cite{Orrit2010} this approach also requires the
emission frequency to fluctuate around one fixed central wavelength, which is in general not the case for many quantum
emitters.
\\
Here, we use an advanced and simplified interferometric setup, combining the advantages of the previous
approaches. The timing resolution is limited only by the single photon counting instrumentation, which can be
100~ps or below. The spectral resolution can almost be as high as for conventional Fourier spectroscopy, while
the central wavelength of the emission does not need to be fixed. Furthermore, the photon detection efficiency
is at least doubled compared to the scheme presented in Ref.~\cite{Sallen2010}. Therefore, our method is
applicable even for moderate integration times.

We use a fixed Mach-Zehnder interferometer as disperse element, converting the
spectral modulation of the ZPL into an intensity modulation which can be measured by correlating the photons from
the two interferometer outputs. Figure~1b shows a sketch of the setup. From the measured photon statistics,
the timescale of the spectral diffusion is derived. In the following we calculate the cross-correlation between
the outputs of the interferometer. We suppress the wavelength index $\lambda$ of the photon number operator
$\hat{n}_{\lambda}(t)$
for simplicity.\\
The time averaged photon detection rate in the left or right output port of the interferometer  is $<\hat{I}(t)_{L/R}>_{t}$, with
\begin{eqnarray}
\hat{I}(t)_{L/R}&=& \eta_{L/R}\, \hat{n}(t) \,m_{L/R}(t),
\end{eqnarray}
where $\eta_{L/R}$ is the overall quantum efficiency in the left and right exit of the interferometer, respectively.
$m_{L/R}(t)$ is the interferometer introduced modulation
\begin{eqnarray}
m_{L/R}(t)&=&1\pm c\, \sin[2\pi x/\lambda(t)],
\end{eqnarray}
where $c$ is the contrast of the interference fringes and $x$ the path length difference of the interferometer
arms. The plus and minus signs correspond to the left and right arm, respectively. The cross-correlation between
the two arms $g^{(2)}_{LR}(\tau) =
\frac{<:\hat{I}_{L}(t)\hat{I}_{R}(t+\tau):>_{t}}{<\hat{I}_{L}(t)>_{t}<\hat{I}_{R}(t)>_{t}}$ reads
\begin{eqnarray}
g^{(2)}_{LR}(\tau)&=&g^{(2)}(\tau) <m_{L}(t)m_{R}(t+\tau)>_{t},
\end{eqnarray}
where $ g^{(2)}(\tau)= \frac{<:\hat{n}(t) \hat{n}(t+\tau):>_{t}}{<\hat{n}(t)>_{t}^{2}}$ is the second order
autocorrelation
function of the bare emitter and $<:...:>_{t}$ denotes normal ordering before time averaging.\\
We assume, that spectral diffusion occurs in form of jumps of the narrow emission line to random positions within a broad
envelope~\cite{Shen2008} and that individual jumps lead to a significant change of the interferometer transmission.
To calculate $<m_{L}(t)m_{R}(t+\tau)>_{t}$ the characteristic function $X(t,\tau)$, which is one if no spectral jump occurred within the time interval $(t,t+\tau)$ and zero else is introduced and terms with and without spectral jump are sperated in the average. By using that $\lambda(t)=\lambda(t+\tau)$ if no jump occurred and $\lambda(t+\tau)=\lambda(t')\neq\lambda(t)$ if a jump occurred, separating $X(t,\tau)$ and identifying the probability that the spectral position remains unchanged after the time $\tau$ as $p(\tau)=<X(t,\tau)>_{t}$ we get:
\begin{eqnarray}
 <m_{L}(t)m_{R}(t+\tau)>_{t}&=&p(\tau)<m_{L}(t)m_{R}(t)>_{t} \nonumber\\
  +\left[1-p(\tau)\right]&\cdot&<m_{L}(t) m_{R}(t')>_{t,t'}.
   \end{eqnarray}
If the interferometer is adjusted in a way that several fringes are within the inhomogeneous width of the ZPL
 the sine-terms average out and  \mbox{$<m_{L}(t) \cdot m_{R}(t')>_{t,t'}=1$}, while \mbox{$<m_{L}(t)m_{R}(t)>_{t}=1-c^{2}/2$}. Therefore
 \begin{eqnarray}
   <m_{L}(t)m_{R}(t+\tau)>_{t}&=&1-c^{2}/2\cdot p(\tau).
   \end{eqnarray}
  Remarkably, here also path length fluctuations on timescales slower than the SD average out and interferometric stability is not required.
Finally, the probability, that the spectral position of the ZPL remains unchanged after some time $\tau$ is
\begin{eqnarray}
 p(\tau) &=& 2/c^{2} \cdot \left(1 - \frac{g^{(2)}_{LR}(\tau)}{g^{(2)}(\tau)}\right).
   \end{eqnarray}
Thus, it is sufficient to measure the autocorrelation $g^{(2)}(\tau)$  of the bare emitter and the
cross-correlation between the two outputs L/R of the interferometer $g^{(2)}_{LR}(\tau)$ to gain knowledge on
the spectral dynamics of the emitter.
\begin{figure}[b]
\centering
 \includegraphics[width=0.6\columnwidth]{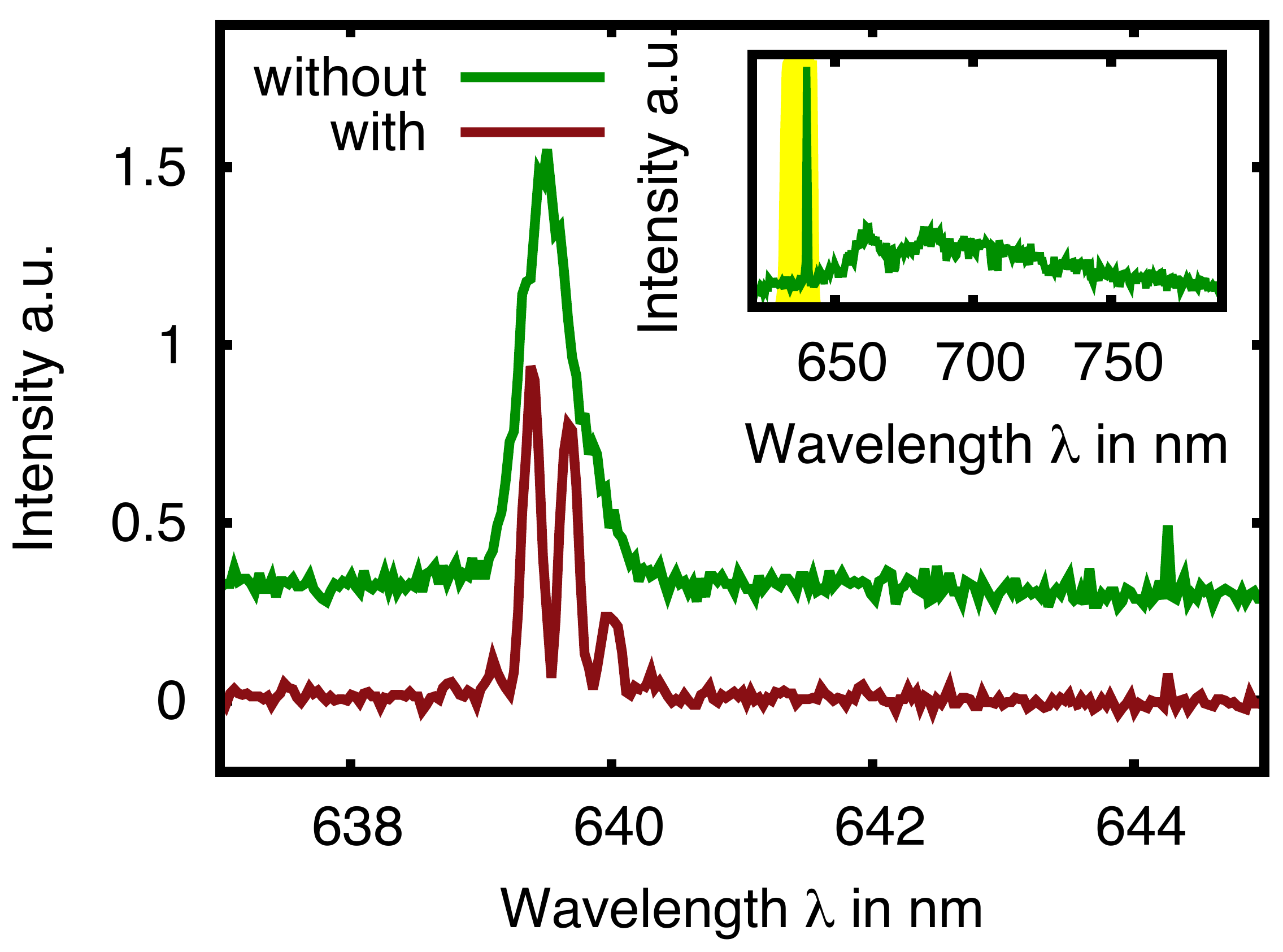}
  \caption{(Color online) Spectra of the ZPL from a typical nanodiamond at a temperature of 5~K with a
  removable bandpass filter centered at the ZPL (637~nm). The green curve
  (upshifted by 0.3 for clarity) is the NV center's fluorescence measured without the interferometer. The lower red curve
  is recorded after the interferometer.
  The inset shows the full NV spectrum with the yellow shaded area indicating the transmission of the bandpass filter.}
  \label{fig:2}
\end{figure}
Nanocrystals milled from high-quality type Ib bulk diamond with a size of 30~nm to
100~nm were spin coated on a solid immersion lens (SIL) made of zirconium dioxide~\cite{Schroeder2011}. The SIL
is placed in a continuous flow He cryostat at about 5~K. Fluorescence is excited and collected in a confocal
configuration (see Fig.~1b) through a commercial NA 0.9 objective lens (\textit{Mitutoyo}) placed inside the
isolation vacuum of the cryostat. The NVs are excited by a green 532 nm laser with a power of several
$\mu$W in front of the vacuum chamber. The single photons emitted by the NV are collected through the
same objective lens and separated by a dichroic mirror from the excitation beam. An additional longpass filter
blocks the excitation laser. In order to suppress the phonon side band, a removable bandpass filter (width 7~nm)
centered at 637~nm is used. The filtered photons are directed into a Mach-Zehnder interferometer and detected by
two avalanche photo diodes (APD) in the two outputs and counted with a time-correlated single photon counting
module (\textit{PicoQuant}). The arm length difference of the interferometer was adjusted to obtain 4 fringes
per nm at 639~nm, the ZPL wavelength
 (see Fig.~2). 
 To check the alignment of the interferometer one of the APDs
was replaced by a 500~mm spectrograph (Acton SpectraPro 500i) with a cooled CCD detector and spectra were
recorded (see Fig.~2).
\begin{figure}[tb]
\centering
 \includegraphics[width=0.49\columnwidth]{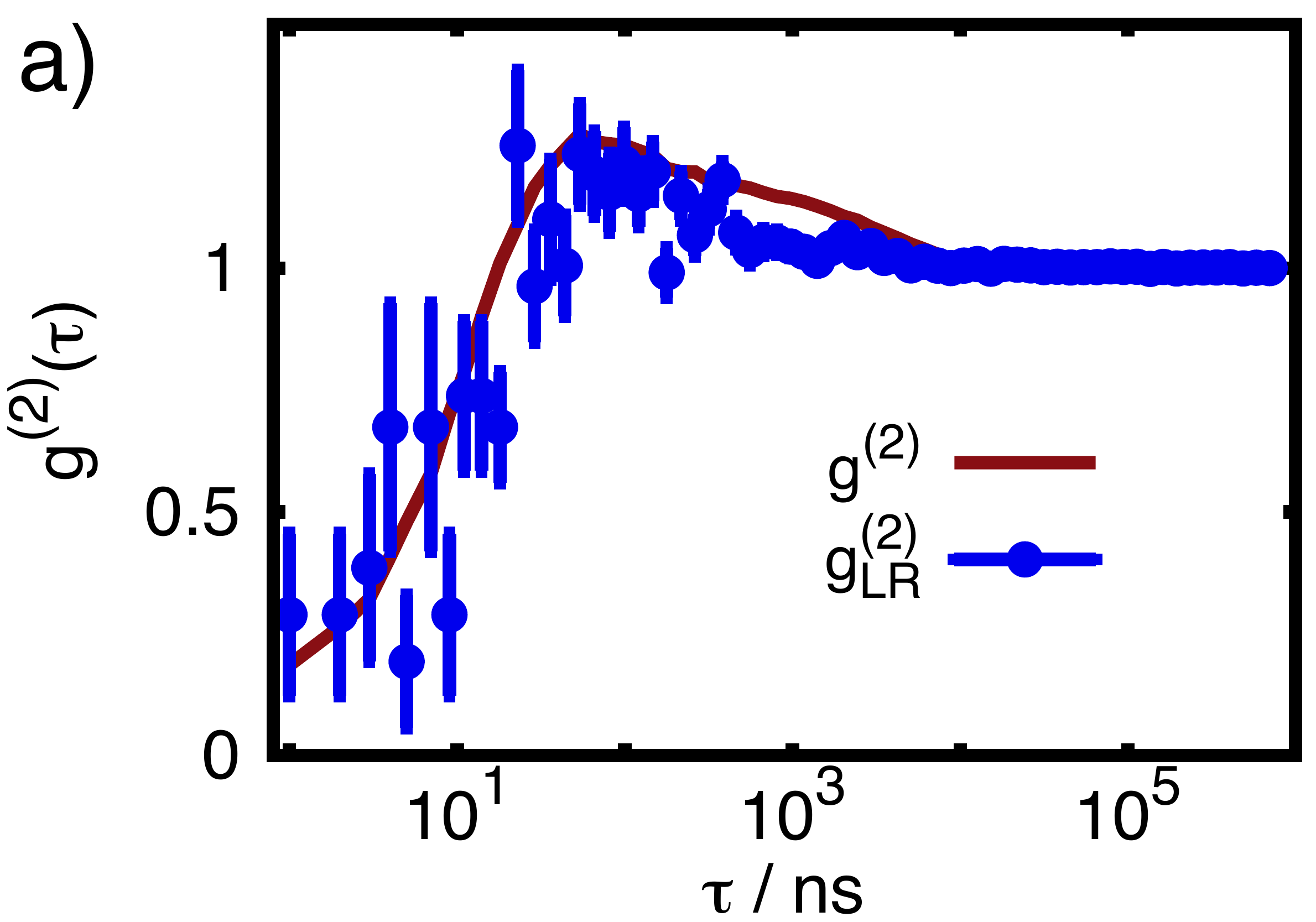}
  \includegraphics[width=0.49\columnwidth]{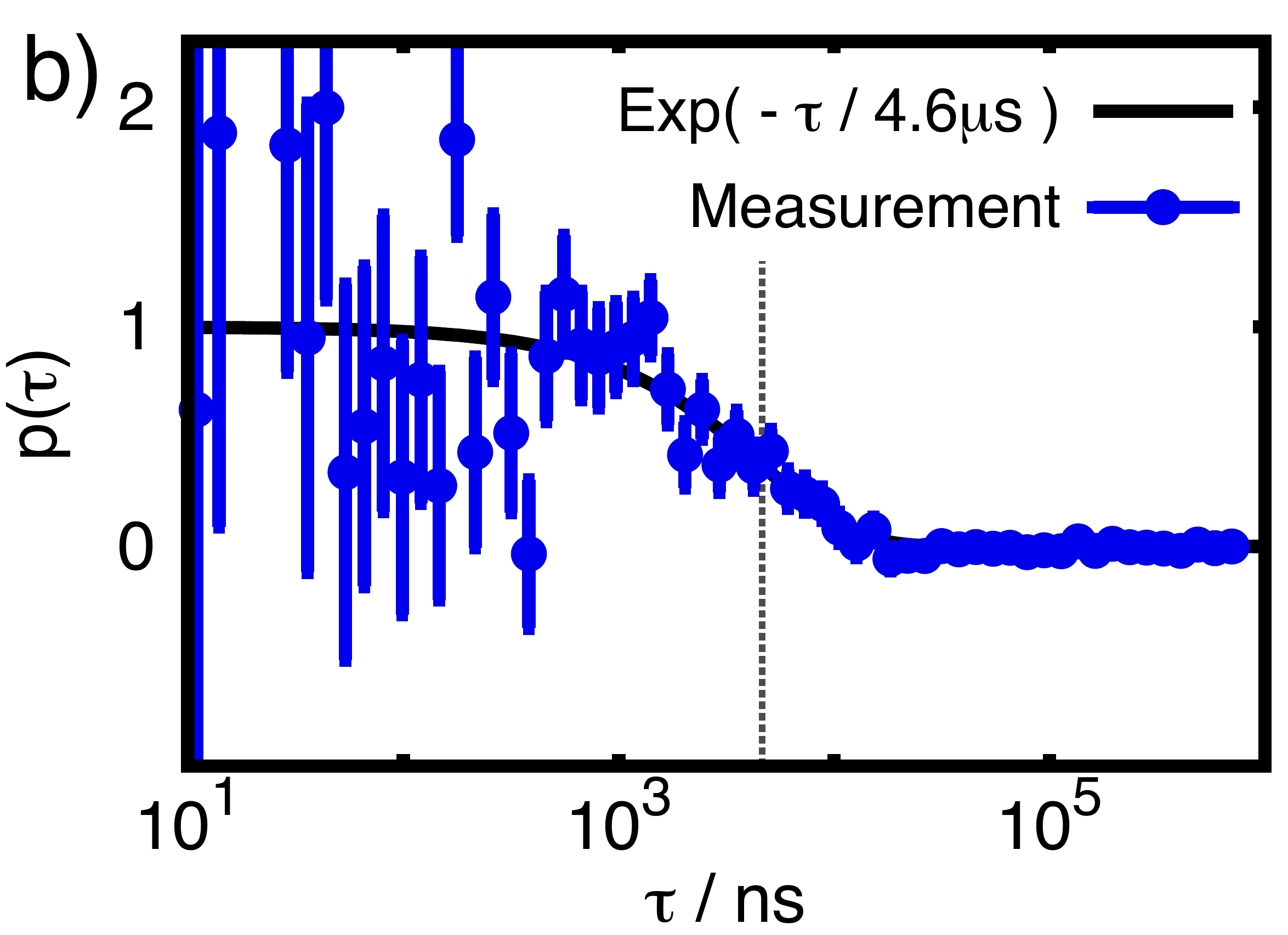}
  \caption{(Color online) a) Measured second-order autocorrelation function $g^{(2)}(\tau)$ (red line) and
  cross-correlation $g^{(2)}_{LR}(\tau)$ (blue dots) of NV-1 at an excitation laser power of 14.1~$\mathrm{\mu W}$. The bin size is increased for longer delays $\tau$, to decrease the error. The error of $g^{(2)}(\tau)$ is in the order of the linewidth. The dip at zero time delay shows the single photon character of the emission.
  b) The probability $p(\tau)$ that the ZPL remains constant on the order of 100~GHz within the time interval $\tau$
  calculated from the data shown in Fig.~3a according to Eq.~(6).  The bin size is increased for longer delays $\tau$, to decrease the error. 
   The black line is an exponential fit to the data showing a spectral diffusion time $\tau_{D}$ of (4.6$\pm$0.6)~$\mu$s, while
  the dashed black line indicates the spectral diffusion time $\tau_D$ (1/e time).
}
  \label{fig:3}
\end{figure}\\
In order to derive the probability $p(\tau)$ that the spectral position remains unchanged after the time $\tau$
according to Eq.~(6) first the bare second-order autocorrelation of the NV emission has to be measured.
In order to do this, the whole fluorescence from the NV including phonon side-bands is sent into the
interferometer. In this case the autocorrelation function $g^{(2)}(\tau)$ measured by the two APDs is
unaffected by the interferometer as proven by an independent measurement. The setup acts as a usual Hanbury
Brown and Twiss setup. Next, the 637~nm bandpass filter is added in front of the interferometer and the
cross-correlation $g^{(2)}_{LR}(\tau)$ between the interferometer arms is measured with an typical integration time of
30~minutes. In this case the interferometer converts the spectral jumps of the narrow ZPL emission into
intensity fluctuations. Care was taken not to change the laser power or adjustment within the total measurement time.
\\
The measured $g^{(2)}$- and $g^{(2)}_{LR}$-functions of NV-1 taken at an excitation laser power of 14.1~$\mathrm{\mu W}$
are shown in Fig.~3a. In order to derive the spectral diffusion probability $p(\tau)$ the data is evaluated
according to Eq.~(6), resulting in the data shown in Fig.~3b. 
Measurements taken with different interferometer adjustments revolving spectral jump widths from 260~GHz to 20~GHz showed, that the assumption of a narrow line jumping in a broad envelope is well justified (see online supplementary).
Obviously, the integration time of 30~min was sufficiently long to resolve the spectral diffusion, but if higher timing resolution is needed the integration time must be increased.
The decrease of the probability $p(\tau)$ fits well to a single exponential decay. This supports the assumption
that spectral diffusion is indeed caused by uncorrelated charge fluctuations. The spectral diffusion
time $\tau_{{D}}$ obtained from the fit is $4.6\pm0.6\,\mu$s corresponding to a spectral diffusion rate
$\gamma_{D}=1/\tau_{D}$ of about 220~kHz, while detection rate at the APDs of photons from the ZPL
transition is only 2.3~kHz.
\\
The achieved contrast $c$ derived from the fit is $35\%$, which is lower than the measured interferometer
contrast of 90\% due to fluorescence background and the two non-degenerate NV dipole
transitions~\cite{Acosta2012}.
\begin{figure}[tb]
\centering
 \includegraphics[width=0.49\columnwidth]{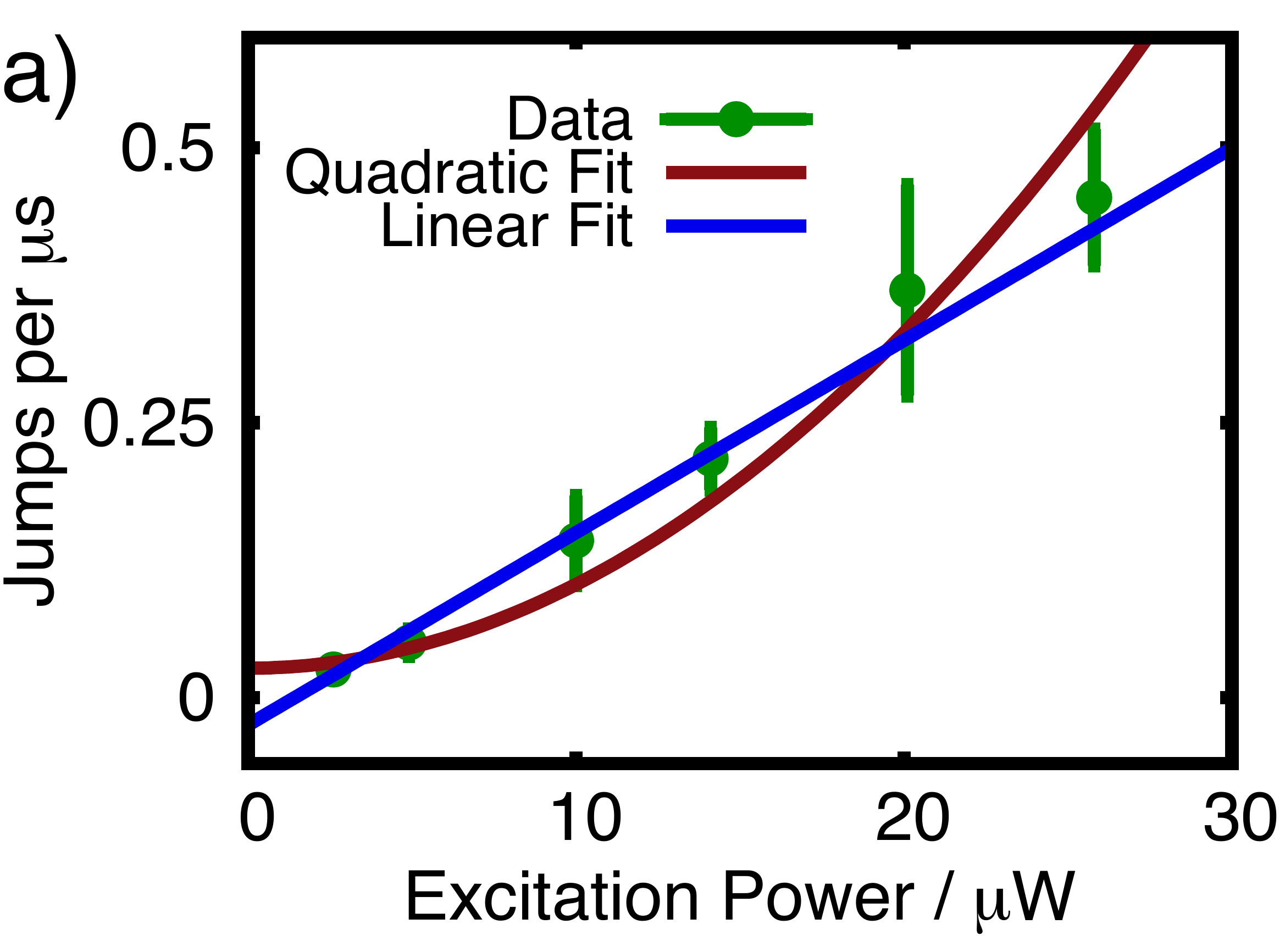}
  \includegraphics[width=0.49\columnwidth]{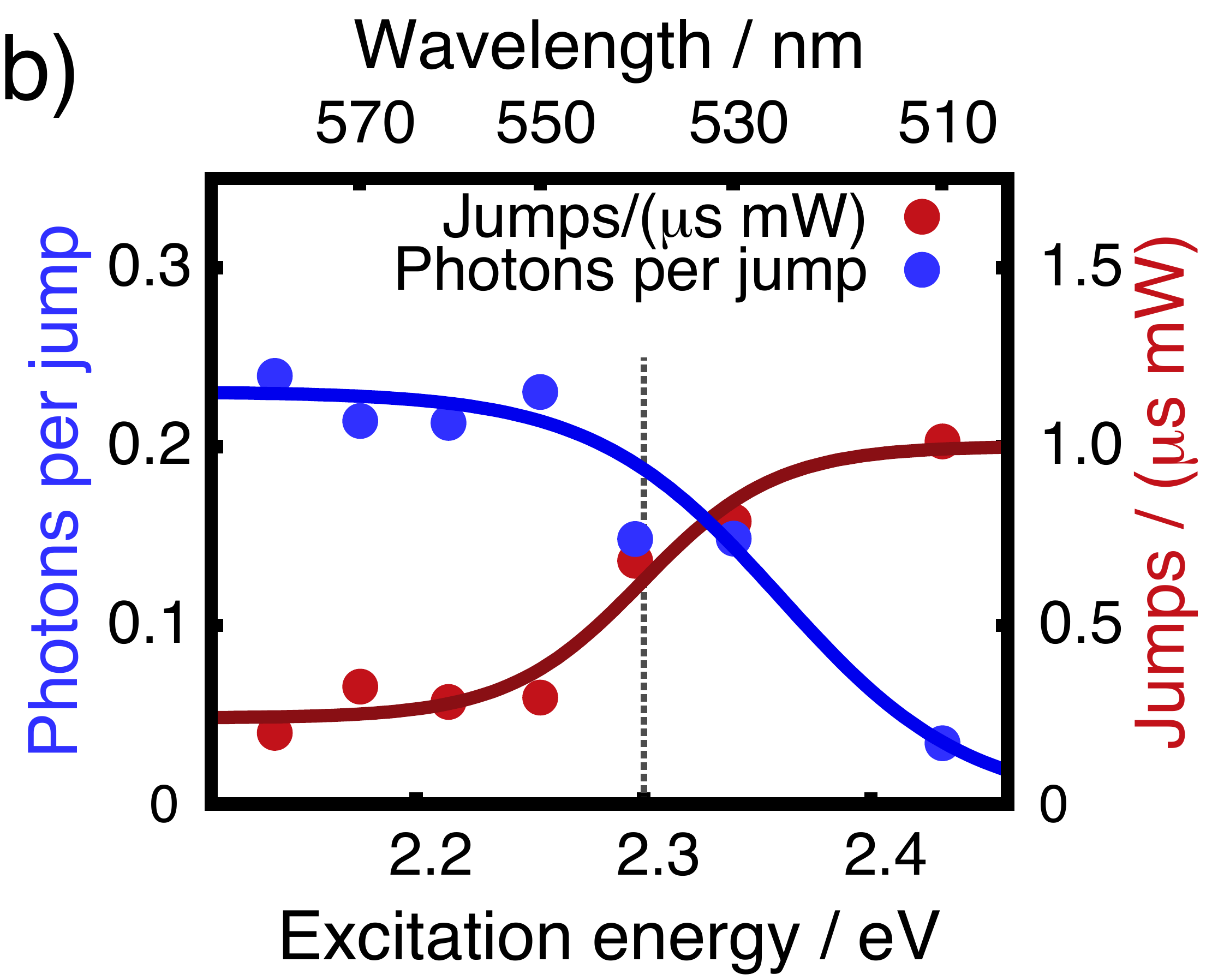}
  \caption{(Color online) a) Dependency of the spectral diffusion rate on the excitation power, measured on NV-1.
  The blue curve is a linear fit to the data with the jump rate at zero excitation power being a free fit parameter.
  The red curve is a quadratic fit to the data. Obviously, the linear function fits better to the data. 
  b) The number of collected photons from the ZPL transition per spectral jump
  as a function of the excitation photon energy, as measured on NV-2. The blue data points were measured at equal photon count rates and
varying excitation powers (left axis). Point size corresponds to the error bar. The blue curve is a guide to the eyes. 
The red points show the spectral jump rate normalized to the excitation power, calculated from the same data set (right axis). 
It is clearly visible, that the spectral diffusion rate increases dramatically above the threshold of about 2.3~eV. 
}
  \label{fig:1}
\end{figure}\\
To get further insight into the origin of the spectral diffusion, we measured the dependency of the spectral
diffusion rate $\gamma_{D}$  on the excitation power on NV-1. Therefore, the above described
measurement was repeated for several laser powers between 3~$\mu$W and 23~$\mu$W, well below the saturation power.
The resulting data (Fig.~4a) show a clear linear dependency, a feature that was also observed on NV-3 where we repeated the measurement (see online supplementary). 
Thereby we can rule out two photon processes like the photo induced charge conversion\cite{Waldherr2011} to be the origin of spectral diffusion.
Remarkably for zero excitation power, the spectral diffusion rate reaches zero within the precision of the fit, although the intersection with the axis is a free fit parameter.
From this we make two conclusions: \textit{First, the excitation with the green laser is
the main cause of spectral diffusion}.
This is consistent with our simple charge trap model. The laser ionizes impurities, providing floating charges which can be trapped in other
charge traps.
\textit{Second, the number of collected photons in the ZPL collected per spectral jump is constant for
excitation far below saturation, i.e. independent of the excitation power}. This number directly evaluates the
quality of the single photon emission in terms of the possible number of subsequent indistinguishable photons
available. This is a key figure of merit for future quantum optics experiments with NV-centers. Obviously, in
the linear regime of excitation power (far below saturation) the quality of single photon emission cannot be
improved by reducing the excitation laser power, as it is often done in experiments with self-assembled quantum
dots~\cite{Robinson2000}.\\
In order to derive a strategy to increase the number of collected photons per spectral jump we investigated its
dependency on the temperature and excitation energy. Within the temperature range of 5~K to 20~K we did not
observe any change in the spectral diffusion rate on \mbox{NV-1}. This is again consistent with the laser being the main cause
of spectral diffusion by ionization of charge traps since their ionization energy exceeds $k_bT$ at this temperature
range.\\
In order to measure the influence of the excitation energy, we replaced the 532~nm cw laser with a pulsed super
continuum source (\textit{NKT Photonics}) with exchangeable 10~nm broad bandpass filters. While keeping the
count rate in the ZPL constant at ($1.4\pm0.2$)~kcts/s (well below saturation) we measured the spectral diffusion rate of NV-2 for several
excitation wavelength from 510~nm to 580~nm, corresponding to photon energies between 2.1~eV and 2.4~eV. We
plotted the number of collected photons in the ZPL per spectral jump and the spectral jump rate normalized to the excitation power. 
The measurement in Fig.~4b clearly indicates, that the number of collected photons per spectral jump decreases
and spectral diffusion rate increases, respectively, with increasing excitation energy. Remarkably, there is a
pronounced threshold of about 2.3~eV. This gives evidence for the existence of deep charge traps with
an ionization energy close to 2.3~eV. 
The remaining spectral diffusion for energies below threshold might be attributed to substitutional nitrogen atoms in the diamond nanocrystals forming donor levels which are ionized at 1.7~eV~\cite{Farrer1969,Rosa1999}.
More extensive studies to identify the trap states as well as to generally modify these states in order to reduce spectral diffusion and studies of higher quality diamond with natural or implanted defects will be a subject of future work. Furthermore using two color excitation, the studies can be extended to energies below the ZPL transition to investigate the influence of the nitrogen donors at 1.7~eV.

In conclusion, we presented an interferometric method to measure the spectral diffusion of the ZPL of single
nitrogen vacancy centers in nanodiamonds. Our method works even if the spectral diffusion rate is orders of magnitudes higher
than the photon detection rate and it is applicable for moderate integration times. Our systematic studies show
that in milled type Ib nanodiamonds the jump rate is 1-2 orders of magnitude higher than the photon
collection rate. Furthermore, the jump rate per photon is independent of the excitation power and NV
temperature, but depends strongly on the excitation wavelength. Our results support the assumption that the main
source of spectral diffusion stems from charge traps. A promising strategy to maximize the number of subsequent
single photons from a nanodiamond is to use relatively high excitation powers, but clearly below the saturation
level and to chose a proper low excitation wavelength. Actively enhancing the optical transition
strength~\cite{Wolters2011,Benson2011} or collection rate is also required, possibly to enable active line
stabilization schemes, as reported for NV centers in bulk diamond~\cite{Acosta2012}. Finally, surface treatment
accompanied by ultra-fast spectral diffusion studies should be extended to learn more about the dynamics of trap
states on the diamond surface. 

This work was supported by the DFG (BE2224/9 and FOR 1493 \textit{Quantum Optics in Diamond}). We thank the
groups of J.~Wrachtrup and F.~Jelezko for providing the diamond nanocrystals. J.~Wolters acknowledges funding by
the state of Berlin (Elsa-Neumann).

\clearpage
\section{Supplementary information}
\begin{figure}[thb]
\centering
 \includegraphics[width=0.4\columnwidth]{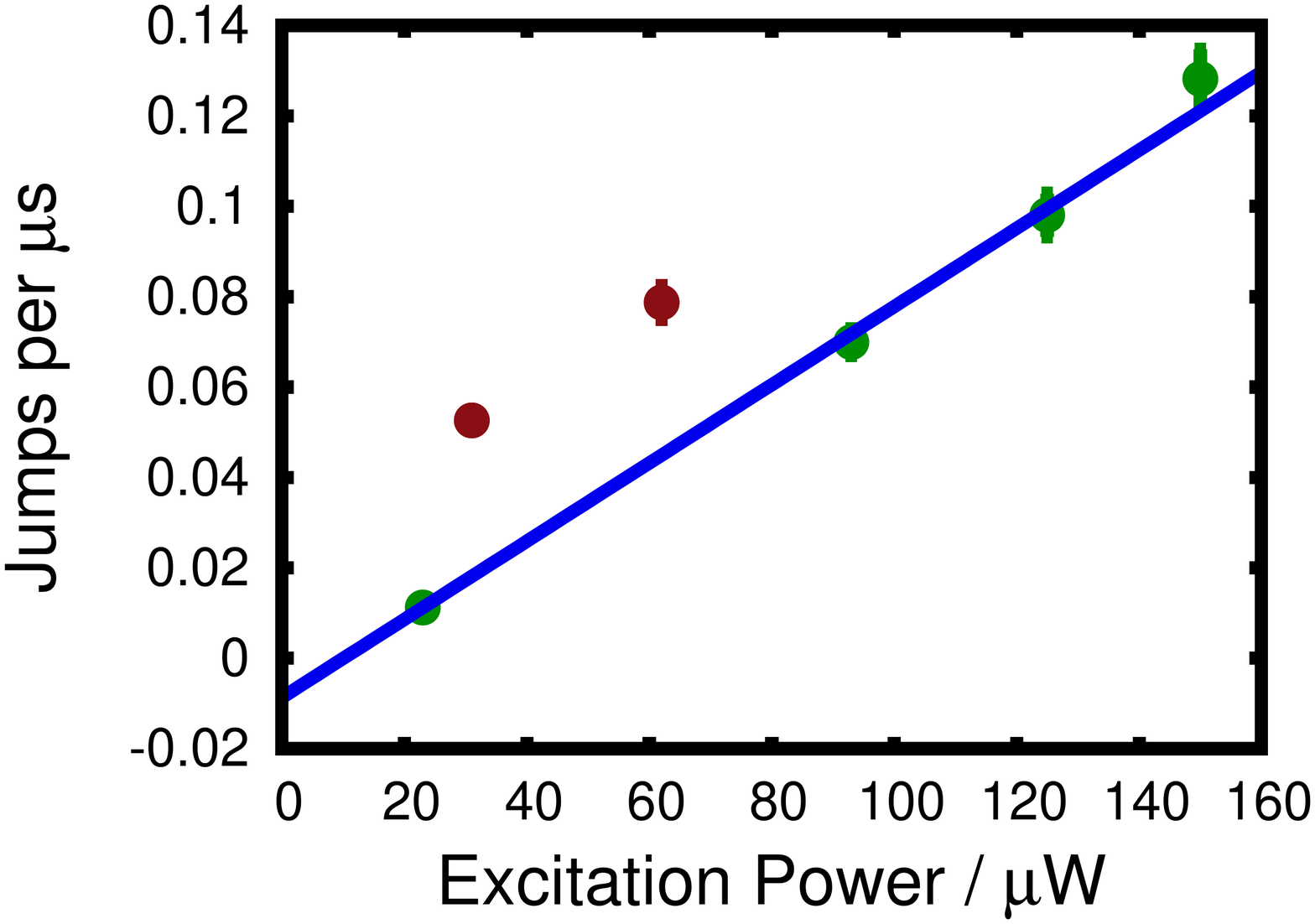}
 \includegraphics[width=0.4\columnwidth]{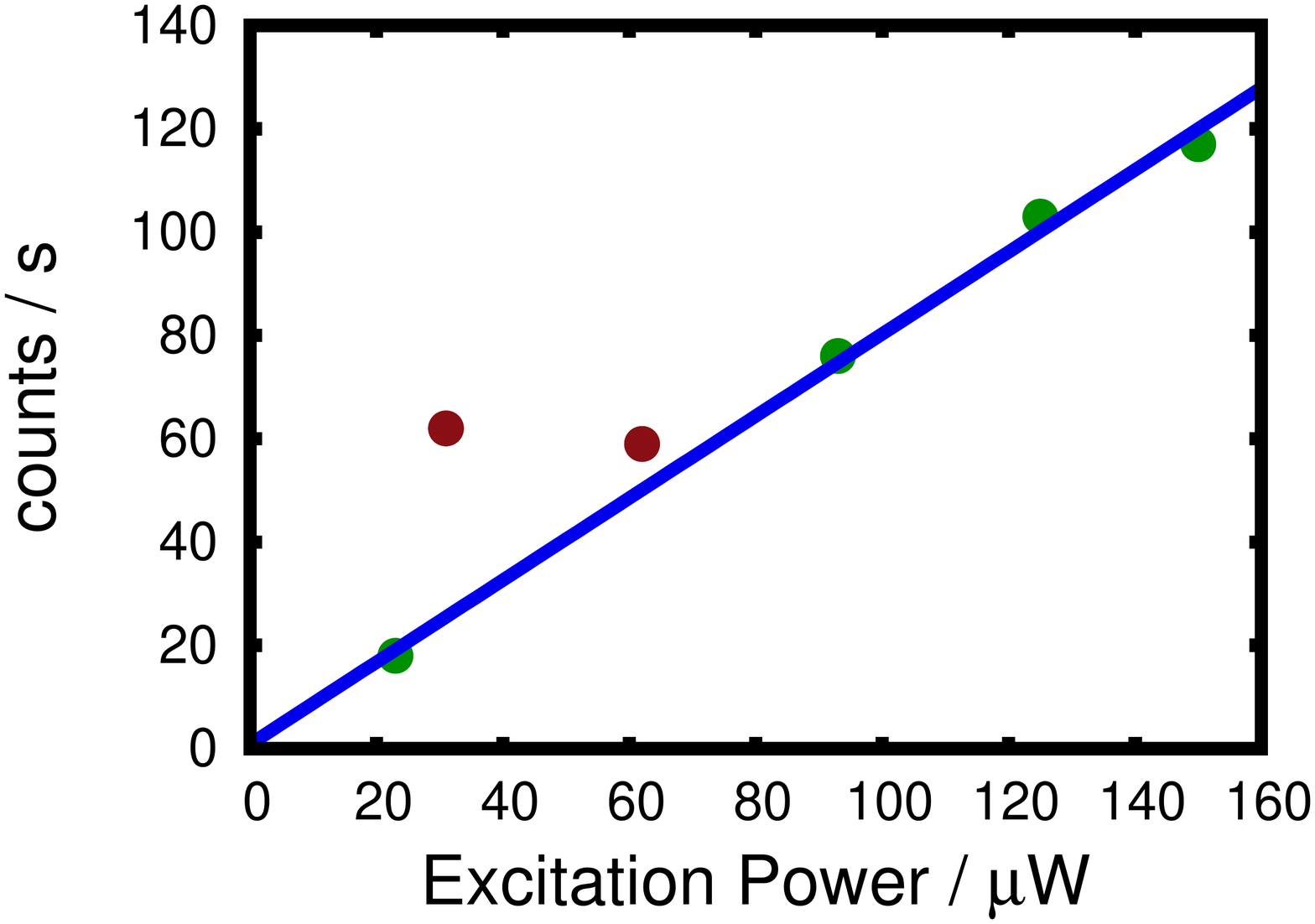}
  \caption{(a): The spectral diffusion rate as a function of excitation power measured on NV-3. The blue curve is a linear fir to the data (green points). The red measurement points were not used for evaluation, as the count rate significantly changed during the measurement. (b):  The single photon count rate as a function of excitation power, measured simultaneously with (a). The red points correspond to measurements, where the count rate was increased due to ablation of dirt.}
  \label{fig:2}
\end{figure}
\begin{figure}[htb]
\centering
 \includegraphics[width=0.4\columnwidth]{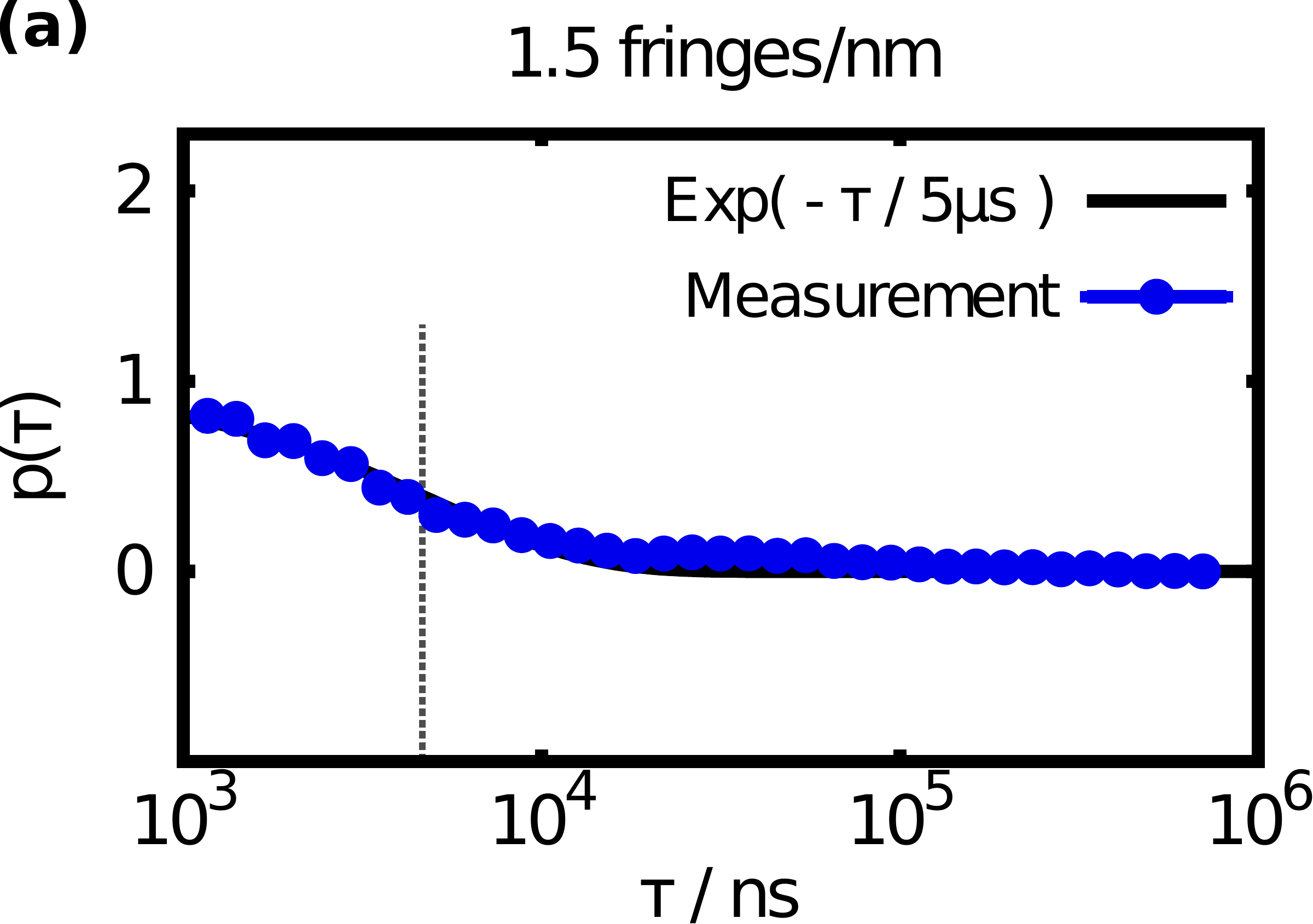}
 \includegraphics[width=0.4\columnwidth]{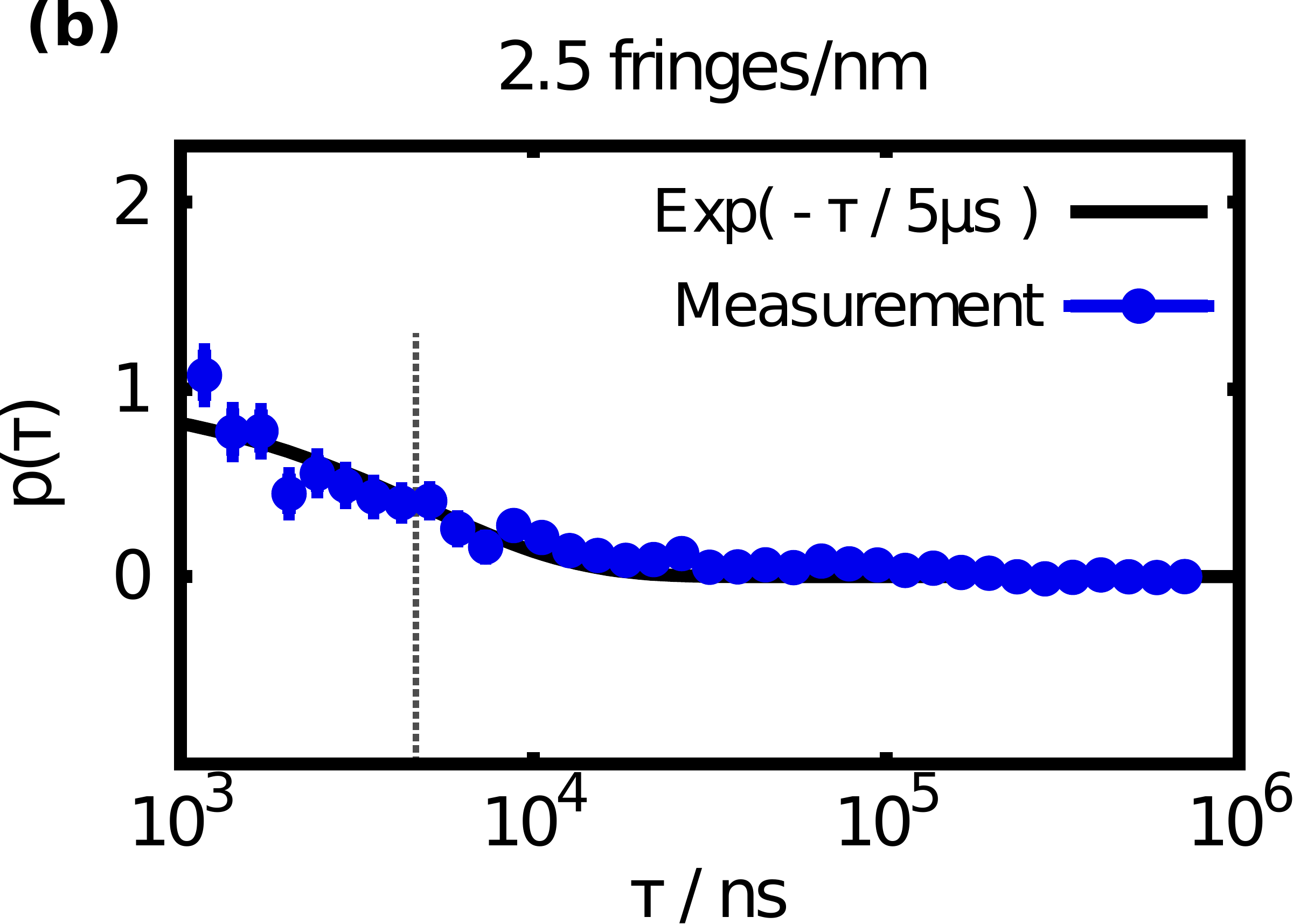}
  \includegraphics[width=0.4\columnwidth]{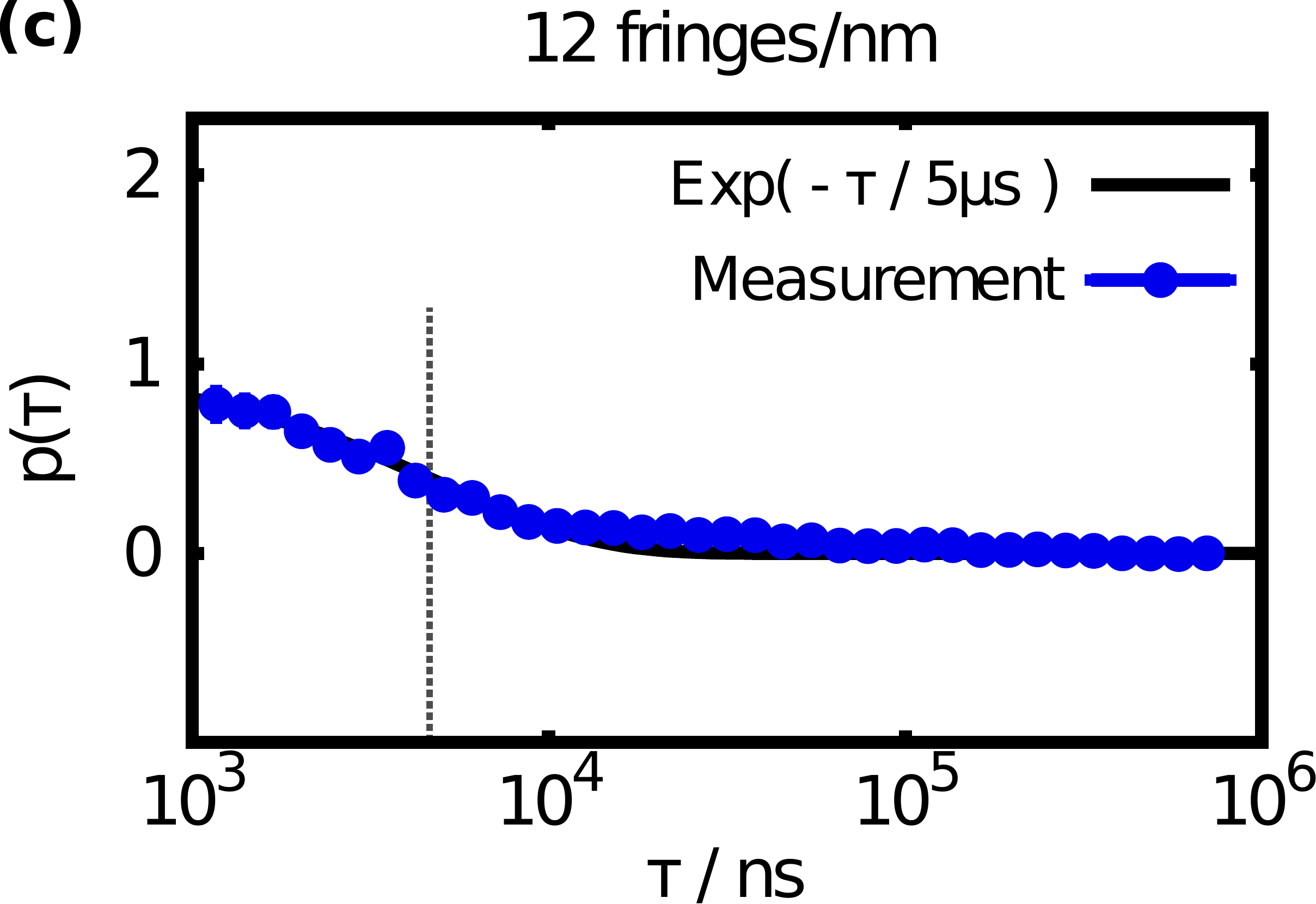}
  \caption{Spectral diffusion time at different interferometer positions measured at NV-4. In (a) spectral jumps larger than  260~GHz are resolved. In (b) the resolution is on the order of 110~GHz, while in (c) jumps larger than 20~GHz are resolved. The spectral diffusion time for all measurements is the same, indicating, that spectral diffusion occurs in form of random jumps of the ZPL with a width below 20~GHz to a new position within the broadened ZPL.}
  \label{fig:2}
\end{figure}

\end{document}